\documentstyle[sprocl]{article}

\def\be{\begin{equation}}
\def\ee{\end{equation}}
\def\bea{\begin{eqnarray}}
\def\eea{\end{eqnarray}}

\begin{document}

\title{ NON-PERTURBATIVE HAMILTONIAN APPROACHES TO STRONG INTERACTION PHYSICS }

\author{J. P. VARY, T.J. FIELDS, J. R. SPENCE}

\address{Department of Physics and Astronomy, Iowa State University \\
Ames, IA 50011, USA\\E-mail: jvary@iastate.edu}

\author{ H.W.L. NAUS}

\address{ Institute for Theoretical Physics, University of Hannover \\
    Appelstr. 2, 30167 Hannover, Germany}

\author{H. J. PIRNER}

\address{ Institute for Theoretical Physics, University of Heidelberg \\
    Philosophenweg 19, 69120 Heidelberg, Germany \\E-mail:
pir@hobbit.mpi-hd.mpg.de }

\author{K. S. GUPTA}

\address{ Saha Institute Of Nuclear Physics, Block - AF, Sector - I,\\
Salt Lake, Calcutta 700064, India\\E-mail: gupta@tnp.saha.ernet.in }


\maketitle\abstracts{
The theory of the strong interactions, Quantum Chromodynamics (QCD), has been
addressed by a variety of non-perturbative techniques over the decades
since its introduction. We have investigated Hamiltonian formulations
with different quantization methods and approximation schemes. In one method,
we utilize light-front coordinates to  investigate the role of
bosonic zero modes in leading to confinement. In another method
we are able to obtain spectra for the mesons and baryons using
constituent quark masses but no phenomenological
confinement.  We survey our principal
accomplishments to date and indicate our future directions.}


\section{Introduction and Motivation}

For strong interaction physics, one might well ask
`Why develop Hamiltonian methods for gauge theories -
after all, does not the lattice gauge method
work well?'

While lattice gauge methods work well for certain
observables, we have a wider range of observables
in mind.  In addition, we are motivated by the
desire to work within a Hamiltonian framework which
we find more physically intuitive.  One final
motivation for us is that we believe there are potential
advantages of using advanced methods from
quantum many-bdoy theory.

On the other hand, we are also quick to acknowledge
many challenges which we face.  For example, little
is known about renormalization and scale
dependence within the Hamiltonian approach to strong
interactions.  We will lose manifest gauge
invariance when we derive a Hamiltonian expressed
only in terms of the independent degrees of
freedom.  Any approximations, such as a truncation,
will usually lead to gauge-dependent results.
Finally, due to our approximations, we may lose
other symmetries respected by the original
Lagrangian.

Our purpose here is to outline our recent progress in
implementing Hamiltonian approaches and to indicate
where we still face major hurdles.

\section{Introduction to Many-Body Theory of $H_{eff}$}

Here we discuss certain aspects of recent developments in
the theory of effective Hamiltonians for quantum many-body systems
which are particularly relevant for our applications to strong
interactions.  Note that these developments
are cast in a form which is independent of the kinematics
of the interacting particles.

The basic framework we adopt has been utilized extensively in nuclear
physics applications.~\cite{zh1}  Our goal is to solve the usual
Hamiltonian eigenvalue problem:
\begin{equation}
H | \Psi_{\alpha}\rangle = E_{\alpha} | \Psi_{\alpha}\rangle	\label{h1}
\end{equation}
for the eigenenergies $E_{\alpha}$ and the eigenstates $| \Psi_{\alpha}\rangle$
of the many-particle system, where $\alpha$ is some label characterizing
the states. But it is impossible to solve this problem in the full
Hilbert space {\bf S} when the number of particles in the system exceeds
3 or 4 because it contains too many degrees of freedom.
Consequently, one wishes to truncate the problem to a smaller
space ${\cal S}$ of dimension $d$, in which it becomes tractable to carry
out the calculation. Now let $| \Phi_{\beta}\rangle$
represent the projections of $d$ of the states $| \Psi_{\beta}\rangle$ into
${\cal S}$. Thus we define the
effective Hamiltonian ${H_{eff}}$ in ${\cal S}$ to satisfy
\begin{equation}
{H_{eff}} |\Phi_{\beta}\rangle = E_{\beta} | \Phi_{\beta}\rangle, \label{h2}
\end{equation}
where the eigenvalues $\{E_{\beta}\}$ are $d$ of the exact
eigenvalues $\{E_{\alpha}\}$ in Eq.(\ref{h1}). Because
the $| \Phi_{\beta}\rangle$ are projections of the $| \Psi_{\alpha}\rangle$,
they are, in general, {\em not} orthogonal.

The question then arises whether an appropriate ${H_{eff}}$ exists
for any given truncation. One can show this to be true by constructing
the biorthogonals to $|\Phi_{\beta}\rangle$, namely,
$|\tilde{\Phi}_{\gamma}\rangle$, which satisfy
$\langle \tilde{\Phi}_{\gamma}|\Phi_{\beta}\rangle
= \delta_{\gamma \beta}$. It then follows that the effective Hamiltonian
${H_{eff}}$ always exists and is of the form
\begin{equation}
{H_{eff}} = \sum_{\beta\in {\cal S}} |\Phi_{\beta}\rangle E_{\beta}
	\langle \tilde{\Phi}_{\beta}| ,          \label{fh}
\end{equation}
which automatically satisfies Eq.(\ref{h2}). As Kirson~\cite{kirson}
has emphasized, the question is {\em not} whether ${H_{eff}}$
exists, but whether it has a {\it simple}
enough form, so as to be {\it useful}.

We use the time-independent-perturbation-theory
approach~\cite{bloch,bran,bk1,eo}
in establishing the connection between ${H_{eff}}$ and $H$.
The basic idea involves
the separation of the Hilbert space
into two parts, using the projection operators $P$
and $Q$, where $P$ defines the truncated or `model' space, defined
by the eigenstates of an unperturbed Hamiltonian $H_0$,
and $Q$ defines the excluded space outside the model space.
The projection operators $P$ and $Q$ define
non-overlapping spaces, so
that $PH_0Q=0$.

In the full Hilbert space, a typical many-body choice for $H$ is of the
form
\begin{equation}
H = \sum_{i=1}^A t_i+ \sum_{i<j}^{A}v_{ij} = T + V = (T+U) + (V-U)
  = H_0 + H_I, 		\label{h4}
\end{equation}
where $U$ is some single-particle or `auxiliary' potential, $H_0=T+U$, and
$H_I= V-U$ is the residual interaction.
Only two-body interactions $v_{ij}$ have been assumed among the A-particles,
but the method can be generalized to many-body forces.

Using the Feshbach projection method~\cite{fesh},
one can explicitly project $H$
into the $P$ and $Q$ spaces and rewrite the $P$ space
equation (omitting the subscript $\alpha$ everywhere) in the form
\begin{equation}
\left[ PHP + PHQ\frac{1}{E-QHQ}QHP\right] P|\Psi\rangle = E P|\Psi\rangle,
					 \label{h5}
\end{equation}
where $P|\Psi\rangle = |\Phi\rangle$. The term in the square brackets defines
the effective Hamiltonian
\begin{equation}
{H_{eff}} = PH_0P + {\cal V}(E),				\label{h7}
\end{equation}
where
\begin{equation}
{\cal V}(E) = PH_IP + PH_IQ\frac{1}{E-QHQ}QH_IP  	\label{h8}
\end{equation}
is the effective interaction.
It should be noted that, in general, ${\cal V}$ is an $A$-particle
operator and the energy $E$ in the denominator
corresponds to one of the exact eigenenergies of the $A$-particle system.

Although ${\cal V}(E)$ is an $A$-particle interaction,
the standard assumption is to approximate it in terms of a
perturbation-theory expansion using two-body interactions.
There are many uncertainties associated with this approach for
conventional nuclear physics applications~\cite{bar1} and
experience gained in those applications may well assist us in
our applications to quantum field theories.

Our approach is to take the $A$-particles as all active (i.e.
we do not assume a passive `core'),and Eq.(\ref{h8})
may be interpreted as
a generalized $A$-particle $G$-matrix equation.
For a one-dimensional model space, the exact solutions for the
eigenvalues are given by
\begin{equation}
E = E_0 + {\cal V}(E), 		\label{h9}
\end{equation}
where $E_0$ is the eigenenergy of the unperturbed Hamiltonian $PH_0P$.
If ${\cal V}(E)$ can be constructed for the $A$-particle system, then
Eq.(\ref{h9}) can be solved diagrammatically or iteratively~\cite{ls,qbox},
to obtain all the eigenenergies of the $A$-particle system whose eigenstates
have non-vanishing projections on the one-dimensional model space.
Examples of these procedures with soluble models prove
instructive~\cite{zvb}.

It is not generally possible to construct the full
$A$-particle $G$ matrix. We approximate it by the two-particle
reaction matrix $G$~\cite{gmatrix} plus higher-order terms.
The two-particle $G$ matrix is simply the infinite sum of
two-particle ladder terms.

The perturbation-theory expansion for ${\cal V}(E)$
is now rewritten as a perturbation series in $G(\omega)$ where
$\omega$ is an arbitrary energy, the `starting' energy, around
which we make an expansion.
In our application to no-core
model spaces, these corrections at the two-particle
level are all of the folded-diagram type.
The remaining corrections generate effective many-body
interactions which are neglected in the applications we
discuss here.

\section{Renormalization of $H_{eff}$}


In this section, we introduce a way of implementing
Wilson renormalization~\cite{wilson} within the
context of
the theory of effective Hamiltonians.~\cite{tjf1}
Our renormalization scheme involves manipulations
at the level of the generalized $G$--matrix. We show
how to calculate the beta function
within this context and exhibit our method
using simple scale--invariant quantum
mechanical systems.

We have argued above that the knowledge of the matrix $G$
allows us to
obtain $H_{eff}$. In
what follows, we shall therefore restrict our attention only to $G$.
For the sake of convenience we choose to work in the momentum
representation where the kinetic energy term in the Hamiltonian
is diagonal.
To introduce the concept of renormalization we shall
focus our attention on the one--particle system.
The matrix elements of $G$ are here given by
\begin{eqnarray}
G_{kk'} &=&  \langle k | PVP | k' \rangle + \nonumber \\
&&+\int \,dp  \,dp' \langle k | PVQ | p \rangle \langle p |
\frac{1}{\omega-QH_0Q}
| p' \rangle \langle p' | QVP | k' \rangle  + \cdots
\label{gmat}
\end{eqnarray}
where we have set $U = 0$ and expanded $QVQ$ out of the
denominator.

Let us suppose that the potential $V$ depends on
a single coupling constant $\mu_0$, which we shall call the bare
coupling constant.
It is clear from Eq. (\ref{gmat})
that the matrix element $G_{kk'}$ will be a function of $\mu_0$.
The expression in Eq. (\ref{gmat}) may, in general, require
regularization
due to the divergence arising from the integral.
The regularization that we choose consists of introducing an
ultraviolet cutoff $\Lambda$.
The matrix element in Eq. (\ref{gmat}) is now a function
of the coupling constant $\mu_0$ and the cutoff
$\Lambda$. At the end of the calculation we must remove the cutoff,
i.e. we must take $\Lambda$ to $\infty$, which, as discussed above,
may in general lead to divergence. One way to avoid the divergence is to
replace the coupling constant $\mu_0$ with a function
of $\Lambda$, which we
denote as $\mu (\Lambda)$,
and then require that
matrix element in Eq. (\ref{gmat}) remain finite and independent
of the cutoff as the cutoff is removed.
In other words, we demand that
\begin{equation}
\lim_{\Lambda \to \infty}\frac{d}{d \Lambda}G_{kk'}(\Lambda, \mu
(\Lambda))=0
\label{der}
\end{equation}
The function $\mu (\Lambda)$ thus plays the role of the renormalized
coupling
constant.

The dependence of the coupling constant on the cutoff is usually
expressed in terms of the beta function, which is defined by
\begin{equation}
\beta (\mu) \equiv \Lambda \frac{d \mu}{d \Lambda}.
\label{beta}
\end{equation}
Within our formalism, Eqs. (\ref{der}) and (\ref{beta}) can be used
to
calculate the beta function.

Note that once Eq. (\ref{der}) is satisfied and $\mu(\Lambda)$ is
determined, then $H_{\rm eff}$, based on $G_{kk'}(\Lambda,
\mu(\Lambda))$, should also be independent of $\Lambda$ as
$\Lambda \to \infty$.  Thus, the complete problem of
renormalization is solved.

We shall now illustrate the method prescribed above in two simple
cases of a Dirac particle in 1 dimension and a Schrodinger particle
in 2
dimensions~\cite{pinsky,huang}.
In both these cases the
interaction potential will be taken as a delta function in position
space :
\begin{equation}
V(x) = -\mu_0\delta^{(n)}(x),
\end{equation}
where $n$ is the dimension of configuration space.
In the momentum space the interaction potential would simply be a
constant, i.e.,
\begin{equation}
V(k) = - \mu_0.
\end{equation}
We will choose $H_0$ to be the pure kinetic operator, and our model
space to consist of all states with
momenta less than $\lambda$.  Thus $Q$ projects onto the momentum
range $[\lambda,\infty]$.

	With the choice of the interaction potential described above, the
series in Eq. (\ref{gmat}) can be summed exactly and is given by
\begin{equation}
G_{kk'}=\frac{-\mu_{0}}{1+\mu_{0}I(\omega)}\delta(k-k'),
\end{equation}
where $I(\omega)$ is given by
\begin{equation}
I(\omega)\equiv \int_{\lambda}^{\infty}\,d^{n}p\frac{1}
{\omega-E_{0}(p)},
\label{I}
\end{equation}
Following the preceeding discussion we now introduce an ultraviolet
cutoff
$\Lambda$.
Replacing $\mu_0$
by the renormalized coupling constant $\mu$ and using Eqs.
(\ref{der}) and
(\ref{beta}),
we obtain the beta function as
\begin{equation}
\beta(\mu) = \mu^2 \Lambda \frac{\partial I}{\partial \Lambda}.
\end{equation}

To obtain the explicit expression for the beta function we need to
evaluate the integral appearing in Eq. (\ref{I}). For the 1
dimensional Dirac
particle we have $n=1$, $E_{0}(p)=p+m$ and
\begin{equation}
I(\omega)\equiv \int_{\lambda}^{\Lambda}\,dp\frac{1}{\omega-(p+m)}
=-\,{\rm
ln}\left(\frac{\omega-(\Lambda+m)}{\omega-(\lambda+m)}\right).
\end{equation}
The corresponding beta function is given by
\begin{equation}
\beta (\mu) = -\mu^2.
\end{equation}

For the Schrodinger particle in 2 dimensions we have
$n=2$ and $E_{0}(p)=p^2/2$ (we set the mass of the particle to unity.)
Proceeding exactly as before, we obtain
\begin{equation}
I(\omega)=-2\pi  \,{\rm ln}\left(\frac{\omega-\Lambda^2}
{\omega-\lambda^2}\right)
\end{equation}
and
\begin{equation}
\beta = -4\pi \mu^2
\end{equation}

Note that the results in both examples above have the desirable
property that the beta function is independent of the model space
cutoff, $\lambda$. The beta functions calculated give rise to
asymptotically free theories and generate the generally accepted
pattern for the flow of the coupling constant for the two examples
described above.

\section{Light-Front QCD Application}


Rather than take a pure light-front quantization approach,
we opt for quantizing on a space-like surface in
`near light-front coordinates' in order
to maintain contact with many of the known results
from the usual equal-time quantization method.~\cite{hwln1}

Here we will sketch the dynamics of the gluonic zero
modes of the color $SU(2)$ QCD Hamiltonian and show how the strong
coupling solutions serve as a basis for
the complete problem.  Only an
external charge density $\rho_m$ is considered here.

\subsection{Color $SU(2)$ QCD Hamiltonian}

The Lagrangian in the near light front coordinate system reads
\begin{equation}
{\cal L} = \frac{1}{2}F^a_{+-}F^a_{+-} + \sum_{i=1,2}
\left( F^a_{+i}F^a_{-i} + \frac{\eta^2}{2}F^a_{+i}F^a_{+i}\right)
-\frac{1}{2}F^a_{12}F^a_{12} - \rho_m^a A_+^a,
\end{equation}
where the color index $a$ is summed from 1 to 3, and the transverse
coordinates are labeled by $i=1,2$.
We will also use the  matrix notation; for example $ A_- = A_-^a \tau^a / 2 $.
The field strength tensor contains the commutator of the gauge fields
and the coupling constant $g$:
\begin{equation}
F_{\mu \nu} = \partial_{\mu} A_{\nu} -\partial_{\nu} A_{\mu}
-i g [A_{\mu}, A_{\nu}] \, .
\end{equation}

The $A^a_+$ coordinates have no momenta conjugate to them.
As a consequence, the
Weyl gauge $A^a_+=0$ is the most natural starting point for
a canonical formulation.
The canonical momenta of the dynamical fields $A_-^a, A_i^a$ are given by
\begin{eqnarray}
\Pi^a_- &=& \frac{\partial{\cal L}}{\partial F^a_{+-}} = F^a_{+-},
\nonumber \\
\Pi^a_i &=& \frac{\partial{\cal L}}{\partial F^a_{+i}} = F^a_{-i}
+ \eta^2 F^a_{+i}.
\end{eqnarray}
>From this, we get the Weyl gauge Hamiltonian density
\begin{equation}
{\cal H}_W = \frac{1}{2} \Pi^a_- \Pi^a_- + \frac{1}{2}F^a_{12}F^a_{12}
+ \frac{1}{2\eta^2}\sum_{i=1,2}\left( \Pi^a_i - F^a_{-i}\right)^2.
\label{Hweyl}
\end{equation}
We choose periodic boundary conditions in
$x^- = \frac{x^0 - x^3}{\sqrt{2}}$
 and $x_{\bot}$
on intervals of size $[0,L]$.  Using the appropriate periodic
delta functions, the quantization is straightforward. However,
the Hamiltonian has to be supplemented by the original Euler--Lagrange
equation for $A_+$ as  constraints on the physical states
\begin{eqnarray}
G^a({x}_{\bot},x^-) | \Phi \rangle &=&
\left( D_-^{ab} \Pi^b_{-} + D_{\bot}^{ab}\Pi^b_{\bot} + g \rho_m^a \right)
 | \Phi \rangle \nonumber \\
&=& \left( D_-^{ab} \Pi^b_{-} + G_{\bot}^{a}\right) | \Phi \rangle
=  0,
\label{Gauss}
\end{eqnarray}
with the covariant derivatives
\begin{eqnarray}
D_{-}^{ab} &=& \partial_{-}\delta^{ab} + g f^{acb} A_{-}^c , \nonumber \\
D_{{\bot}}^{ab} &=& \partial_{{\bot}}\delta^{ab} + g f^{acb} A_{{\bot}}^c ,
\label{Deriv}
\end{eqnarray}
where $f^{acb}$ are the structure constants of $SU(2)$.
These equations are known as Gauss' Law constraints.
Since the Gauss' Law operator commutes with the Hamiltonian
\begin{equation}
[ G^a({x}_{\bot},x^-) ,  H_{W} ]    = 0 ,
\end{equation}
time evolution leaves the system in the space of physical states.
Furthermore, $H_{W}$ is invariant under time independent residual
gauge transformations whose  generator is
closely connected to Gauss' Law~\cite{LNT}.

In order to obtain a Hamiltonian formulated in terms of unconstrained
variables,
we eliminate the $A_-$.
Classically this would correspond to the
light front gauge $A_- =0$. However, this choice is not
compatible with gauge invariance and periodic boundary conditions.
Only the (classical) Coulomb light front gauge ($\partial_{-}A_{-}=0$) is
legitimate.
The reason is that $A_-$ carries information on gauge invariant
quantities, such as the eigenvalues of the spatial Polyakov (Wilson) loop
\begin{equation}
{\cal P}({x}_\bot)=  P \exp\left[ ig\int dx^-A_-(x_{\bot},x^-)
\right] \, ,
\end{equation}
which can be written in terms of a diagonal matrix
$a_-({x}_{\bot})$
\begin{equation}
{\cal P}({x}_\bot)= V \exp\left[ ig L a_-(x_{\bot}) \right] V^{\dagger} \, .
\end{equation}
Thus, we obviously need to keep these `zero modes' $a_-({x}_{\bot})$
as dynamical variables.
In order to eliminate the conjugate
momentum, $\Pi_-$ of $A_-$, by means
of Gauss' Law,  one needs to `invert' the
covariant derivative $D_-$. After an unitary transformation $D_-$
simplifies significantly (compare to Eq. (\ref{Deriv}))
\begin{equation}
D_- \rightarrow d_- = \partial_- -ig \, [a_-,  \;\;\;\;\;\;\;  .
\end{equation}
Now Gauss' Law can be readily resolved: in the physical space
one can make the replacement
\begin{equation}
\Pi_{-}({x}_{\bot},x_{-}) \rightarrow p_{-}({x}_{\bot}) - \left (
d_-^{-1} \right ) G_{\bot}({x}_{\bot},y^-) .
\end{equation}
The zero mode  operator $p_-({x}_{\bot})$
is the conjugate momentum
to $a_{- }(x_{\bot})$. It has eigenvalue zero with respect to $d_-$,
i.e. $d_- p_- =0$,
and is therefore not constrained.

The appearance of the zero modes
also implies residual Gauss' Law constraints. In the space of
transformed physical states $|\chi \rangle$, they read
\begin{equation}
\int dx^- G_{\bot}^3 \, |\chi \rangle =
\int dx^- \left ( D_{\bot}^{3b} \Pi_{\bot}^b + g\rho_m^3 \right )
|\chi \rangle =0.
\label{RG}
\end{equation}
These two--dimensional constraints can be handled in full analogy to
QED, since they correspond to the diagonal part of color space.
This further gauge fixing in the $SU(2)$ 3--direction is done via
another gauge fixing transformation, which leads
to the Coulomb gauge representation in the transverse plane for the neutral
fields.  In other words, we eliminate the color neutral,
$x^-$--independent, two--dimensional longitudinal gauge fields
\begin{equation}
{a}_{\bot}^{\ell}(x_{\bot}) = \frac{1}{L} \int dy^- dy_{\bot}
d(x_{\bot}-y_{\bot}) \nabla_{\bot}
\left( \nabla_{\bot} \cdot A_{\bot}^3({y}_{\bot}, y_-) \right)
 \frac{\tau^3}{2} \,.
\label{SUB}
\end{equation}
Here we use the periodic Greens function
of the two dimensional Laplace operator
\begin{equation}
d({z}_{\bot}) = - \frac{1}{L^2} \sum_{\vec{n} \neq \vec{0} }
\frac{1}{p_n^2} e^{i p_n z_{\bot}}  \ ,
\ \ \ \ p_n = \frac{2 \pi}{L} \vec{n} \ ,
\label{Green2}
\end{equation}
where $\vec{n} = (n_1,n_2)$ and $n_1,n_2$ are integers.
The conjugate momenta of these fields,
$p_{\bot}^{\ell} (x_{\bot})$,
are defined analogously.
Resolution of the residual Gauss' Law allows one to replace them,
in the sector of the transformed physical space $| \Phi' \rangle$,
by the  neutral chromo-electric field\\
\begin{eqnarray}
 e_{\bot} (x_{\bot}) &=& g \nabla_{\bot}\int dy^{-}dy_{\bot}
d(x_{\bot} - y_{\bot})
\{f^{3ab} A^a_{\bot}(y_{\bot}, y^-)  \Pi^b_{\bot}(y_{\bot}, y^-)
 \hspace{2cm} \nonumber\\
 & & + \rho^3_m(y_{\bot}, y^-) \}
    \frac{\tau^3}{2} \, .
\end{eqnarray}
It is convenient to introduce
the unconstrained gauge fields and their conjugate momenta:
\begin{eqnarray}
A_{\bot}'(x_{\bot}, x^-) &=&
A_{\bot}(x_{\bot},x^{-}) - a_{\bot}^{\ell}(x_{\bot}),
 \nonumber \\
\Pi_{\bot}'(x_{\bot}, x^-) &=& \Pi_{\bot}(x_{\bot}, x^-) -
\frac{1}{L} p_{\bot}^{\ell}(x_{\bot}).
\label{subt}
\end{eqnarray}
These  relations turn out to be  important for
neutral gluon exchange; recall that the subtracted fields
are diagonal in color space.
Note that the physical degrees of freedom $A_{\bot}'$ and
$\Pi_{\bot}'$ still contain $(x_{\bot}, x^-)$--independent,
color neutral, modes. Therefore, there is a
remnant of the local Gauss' Law constraints -- the global
condition
\begin{equation}
Q^3 |\Phi ' \rangle =
\int dy^{-}dy_{\bot}
\left \{ f^{3ab} A^a_{\bot}(y_{\bot}, y^-) \Pi^b_{\bot}(y_{\bot}, y^-)
+ \rho^3_m(y_{\bot}, y^-)  \right \}
 |\Phi ' \rangle =0 \,.
\label{Gaussgl}
\end{equation}
Its physical meaning is
that the neutral component of the total color
charge, including external matter as well as gluonic contributions, must
vanish in the sector of physical states.

The final Hamiltonian density in the physical sector explicitly reads
\begin{eqnarray}
{\cal H} & = & \mbox{tr} \left[  \partial_1 A'_2 -\partial_2 A'_1
-ig [A'_1, A'_2]  \right]^2
 +  \frac{1}{\eta^2} \mbox{tr} \left[
\Pi'_{\bot}-\left(\partial_-A'_{\bot}-
ig[a_-,A'_{\bot}]\right)\right]^2 \nonumber \\
& + &  \frac{1}{\eta^2} \mbox{tr} \left[ \frac{1}{L} e_{\bot}
-
\nabla_{\bot} a_-\right]^2
 +  \frac{1}{2 L^{2}}  p_{-}^
{3 \, \dagger}(x_{\bot})
p^3_{-}(x_{\bot}) \\
& + &   \frac{1}{L^{2}}\int_{0}^{L} dz^{-} \int_{0}^{L} dy^{-}
\sum_{p,q,n}\,^{'}\frac{ G'_{\bot qp}(x_{\bot}, z^{-})
G'_{\bot pq}(x_{\bot},y^{-})}{\left[ \frac{2\pi n}{L} +
g(a_{-q}(x_{\bot})- a_{-p}(x_{\bot})) \right]^{2}}
e^{i2\pi n(z^{-}-y^{-})/L} \ , \nonumber
\label{Hamil}
\end{eqnarray}
where $p$ and $q$ are matrix labels for rows and columns,
 $a_{-q}=(a_-)_{qq}$ and the prime indicates that the summation is
restricted to $n \neq 0$ if $p=q$.
The operator $G'_{\perp}\left(x_{\bot}, x^-\right)$ is defined as
\begin{equation}
G'_{\perp}=
\nabla_{\perp}\Pi'_{\perp}
+ gf^{abc}\frac{\tau^{a}}{2}A'\,^{b}_{\perp}
\left({\Pi'}\,^{c}_{\perp} -
\frac{1}{L}{e}\,^{c}_{\bot}\right)
+ g\rho_{m}   \,.
\label{Gop}
\end{equation}

\subsection{Zero Mode Dynamics}

The principal advantage of an exact light front formulation is the apparent
triviality of the ground state which simplifies
calculations of the hadron spectrum.
The light front vacuum, however,  is not guaranteed to be trivial in the zero
mode sector.

As can be seen
from the dispersion relation for massless particles on the light front,
$p_+ = \frac{p^2_{\bot}}{2p_-},$
soft momentum modes become high energy states.
In this way, high energy physics
becomes tied to long range physics, contrary to the equal time
formulation.
This physics appears in deep inelastic
scattering at small scaling variable and is related to the long
distance features of the proton.
We will focus on the zero  mode sector in order to
try to acquire some insight into its dynamics.

>From the comparison of abelian and non-abelian
theories, striking differences show up in the zero mode
sector.
Recently, in the equal time formalism, the zero mode sector in
QCD has been
claimed to be relevant for the confinement phenomenon \cite{LMT}.
On the level of approximations and restrictions followed below,
the formal differences between light front and equal time
approach are rather small and, consequently,  results and methods are similar.

In this  work
we do not restrict ourselves to the strong coupling
approximation. We will, however, start with the strongly coupled
theory to define our set of basis functions.
We will restrict ourselves to
pure gluonic $SU(2)$ Yang-Mills theory. It already has the typical non-abelian
features such as the Coulomb term which explicitly contains the zero
modes in the denominator and the non-standard kinetic energy for
the zero modes.

The zero mode degrees of freedom couple to the three--dimensional
gluon fields  via
the second  term in ${\cal H}$ shifting the longitudinal
momenta of the transverse
gluon fields (Eq. (\ref{Hamil})). We neglect these couplings and
consider the pure zero mode Hamiltonian
\begin{equation}
h = \int d^2x \; \left[
 \frac{1}{2 L}  p^{3\,\dagger}_-({x}_{\bot}) p^{3}_-({x}_{\bot})
+ \frac{L}{2\eta^2} (\nabla_{\bot} a^3_-({x}_{\bot}))^2\right].
\label{Hamil3}
\end{equation}
We recognize `electric' and `magnetic' contributions in $h$, the zero mode
Hamiltonian -- the first and second term, respectively.
The light front variables mix the ordinary spatial and time
variables so
the labeling above is to be understood in analogy with the equal time
Hamiltonian.
Even at this level of approximation, this zero mode
Hamiltonian differs from the corresponding one in QED. The reason
is the hermiticity defect of the canonical momentum:
$p_-^{\dagger} \neq  p_-$.

We now omit the color index and introduce dimensionless variables
\begin{equation}
\varphi({x}_{\bot}) = \frac{gL}{2}a_-({x}_{\bot}) \; .
\end{equation}
In the Schr\"odinger representation we then obtain
\begin{equation}
h =  \int d^2x_{\bot} \; \left[
 -\frac{g^2 L}{8} \frac{1}{J(\varphi({x}_{\bot}))}
\frac{\delta}{\delta \varphi({x}_{\bot})}
 J(\varphi({x}_{\bot})) \frac{\delta}{\delta
\varphi({x}_{\bot})}
+ \frac{2}{\eta^2 g^2 L} (\nabla_{\bot} \varphi({x}_{\bot}))^2\right] \; ,
\label{Hamil5}
\end{equation}
where $J(\varphi)$ is the Jacobian
and equals the Haar measure of $SU(2)$
\begin{equation}
 J(\varphi({x}_{\bot})) = \sin^2 (\varphi({x}_{\bot})).
\end{equation}
The Jacobian is connected to the hermiticity defect of $p_-$;
which stems from the gauge fixing procedure taking into account the
curvilinear coordinates. The measure also appears in the
integration volume element for calculating matrix elements.
As in earlier approaches~\cite{Bronz}, $\varphi$ will be
treated as a compact variable, $ 0 \le \varphi < \pi$.

At this stage it is necessary to appeal to the physics of the infinite
momentum frame to factorize the reduced true energy $h_{\rm red}$
and the
Lorentz boost factor $\frac{\gamma}{\sqrt{2}} = \frac{1}{2 \eta}$, since
essentially $h$ is a light front energy, and it is well known how
it behaves under a Lorentz transformation.
Thus we rewrite
$ h_{\rm red} = 2 \eta h \, .$
Since the integral over
transverse coordinates can contain arbitrarily small wavelengths,
we regularize $h_{\rm red}$
by introducing a lattice to evaluate the transverse
integral. The lattice vector $\vec b$ numbers the lattice sites,
and $\vec \varepsilon_{1}$ and $\vec \varepsilon_{2}$
are the two unit vectors on the two--dimensional lattice.
In order to have standard commutation relations on the lattice
the derivative on the lattice becomes
$\frac{\delta}{\delta\varphi_{\vec b}}=\frac{\delta}{\delta\varphi
(x_{\bot}) }a^2$.
We further explicitly pull out the
dependence on the lattice cutoff by defining a new reduced Hamiltonian
$\hat h_{\rm red} = a h_{\rm red} = 2 \eta a h$ and
substituting $\eta=\frac{1}{\sqrt 2} a M$, where
$M$ is a typical hadronic mass~\cite{tjthesis}. This yields
\begin{equation}
\hat h_{\rm red}=\sum_{\vec b}\hat h^{\rm e}_{\vec b}+\sum_{\vec b}
\hat h^{\rm m}_{\vec b},
\label{hred}
\end{equation}
with the electric contribution
\begin{equation}
\hat h^{\rm e}_{\vec b}=-g^{2}_{\rm eff}
\frac{1}{J}
\frac{\delta}{\delta\varphi_{\vec b}} J\frac{\delta}{\delta
\varphi_{\vec b}},\end{equation}
and the magnetic term
\begin{equation}
\hat h^{\rm m}_{\vec b}=\frac{1}{g^{2}_{\rm eff}} \sum_{\vec{\varepsilon}}
(\varphi_{\vec b}-\varphi_{\vec b+\vec\varepsilon})^2.\end{equation}
Since the effective coupling constant,
\begin{equation}
g^{2}_{\rm eff}= \frac{g^2LM}{4\sqrt{2}} \, ,
\end{equation}
contains the large factor $LM$, the product of lattice size in the
longitudinal direction and the hadron mass, a strong coupling
approach seems to be a good starting point.
Note that we avoid introducing `radial wave functions'
or effective potentials as others have done ~\cite{LNT,LMT}.
For each lattice site $\vec b$, $\hat h^{\rm e}_{\vec b}$
(the kinetic energy) has the Gegenbauer
polynomials $C_{n_{\vec{b}}}(\varphi_{\vec b})$ for eigenfunctions:
\begin{equation}
\hat h^{\rm e}_{\vec b} C_{n_{\vec{b}}}(\varphi_{\vec b})=
g^{2}_{\rm eff}
n_{\vec b}(n_{\vec b}+2)
C_{n_{\vec{b}}}(\varphi_{\vec b}) , \end{equation}
with
\begin{equation}
C_{n_{\vec{b}}}(\varphi_{\vec b})=\sqrt{\frac{2}{\pi}}\left\{\frac{
\sin((n_{\vec b}+1)\varphi_{\vec b})}
{\sin\varphi_{\vec b}}\right\} ,
\label{Basis}
\end{equation}
and
\begin{equation}
\int^\pi_0  J(\varphi) C_{n}(\varphi) C_{m}(\varphi_)d\varphi
=\delta_{n m} \; .
\end{equation}
The strong coupling  wave functions of the full
transverse lattice are product states
characterized by a set of quantum numbers $\{n\} =
\{n_{\vec{b}}\}$,
\begin{equation}\label{23}
\Psi_{\{n\}}(\varphi)=\prod_{\vec b} C_{n_{\vec{b}}}(\varphi_{\vec b}) \, .
\end{equation}
These functions form a complete and orthonormal basis for the zero
mode sector.
They satisfy the energy eigenvalue equation
\begin{equation}
\sum_{\vec b}\hat h^{\rm e}_{\vec b}\Psi_{\{n\}}(\varphi)=
g^{2}_{\rm eff}\sum_{\vec b} n_{\vec b}(n_{\vec
b}+2)\Psi_{\{n\}}(\varphi).\end{equation}
The ground state in this limit corresponds to all
$n_{\vec{b}} = 0$ -- a constant wave function
\begin{equation}
\Psi_{\{0\}}(\varphi)=
\prod_{\vec{b} } \sqrt{\frac{2}{\pi}} \; ,
\end{equation}
and the ground
state energy is zero
\begin{equation}
E_{0}=0.
\end{equation}
The first excited energy level is $N_{\bot}^2$-fold degenerate - an excitation
at a single lattice point
\begin{equation}
\Psi_{\{1\}}(\varphi)= \sqrt{\frac{2}{\pi}}
\frac{\sin\left( 2\varphi_{\vec{b}}\right)} {\sin\varphi_{\vec{b}}}
\prod_{\vec{b}' \ne \vec{b}} \sqrt{\frac{2}{\pi}} \; .
\end{equation}
In strong coupling this level is separated by a large amount from the ground
state energy
\begin{equation}
E_{1}=3 g^2_{\rm eff} .
\end{equation}
So far our results are equivalent to those reported by others~\cite{LMT} to
within
re--definitions of wave functions and integration measures.
Weak coupling variational
solutions for the
full $SU(2)$ lattice Hamiltonian have also been given~\cite{Bronz}.
Furthermore,
studies in (1+1)--dimensional Yang-Mills theory \cite{Shaba}
give formal extensions to construct wave functions  for
$SU(N)$ gauge theories.

The magnetic term of the Hamiltonian  couples nearest neighbor
lattice points. In the strong coupling limit its contribution may be
obtained perturbatively (as it has the coefficient $1/g_{\rm eff}^2$)
by evaluating it with the basis function of the ground state.
The result of this is
\begin{equation}
\langle\Psi_{\{0\}} |\sum_{\vec b} h^{\rm m}_{\vec b}|\Psi_{\{0\}}\rangle
=\frac{1}{g^2_{\rm eff}}
\left(\frac{\pi^2}{6}-1\right)\cdot(2N_\bot^2) \, .\end{equation}
Since this energy is proportional to  $ N^2_{\bot} = (L/a)^2$ and
$g_{\rm eff}^2$ grows linearly with $L$,
this part of the zero mode dynamics represents a negligible surface effect
for the three--dimensional system in the strong coupling approximation.

Next,  we discuss
the weak coupling limit $g^2\to 0$. In this case
we can simplify the kinetic
term of the
Hamiltonian by defining new variables $\alpha_{\vec b}$:
\begin{equation}
\alpha_{\vec b}=\frac{\varphi_{\vec b}} {\kappa g} \, ,
\end{equation}
with
$ 8 \kappa^2= \sqrt2 L M \, .$
After expanding in $g$, the reduced Hamiltonian becomes
\begin{eqnarray}
\hat h_{\rm red}&=&
\sum_{\vec b} \left \{
-\left(\frac{\partial^2}{\partial \alpha_{\vec b}^2}
+\frac{2}{\alpha_{\vec b} }\frac{\partial}{\partial \alpha_{\vec b}
}\right)
+\sum_{\vec\varepsilon}(\alpha_{\vec b}-\alpha_{\vec b + \vec \epsilon}
)^2 \right \} .
\label{spin}
\end{eqnarray}
Going over to
Fourier momentum representation,
\begin{equation}
\alpha_{\vec b}=\sum_{\vec k} e^{i\vec k\vec b}R_{\vec k} \, ,
\end{equation}
with $k_i= 2\pi n_{i} / N_{\bot} a $ and $ n_{i}=0,\pm 1,\pm 2,\ldots$
we have
\begin{equation}
\hat h_{\rm red}=\sum_{\vec k}\left \{
- \left(\frac{\partial^2}{\partial
R^2_{\vec k}}+\frac{2}{R_{\vec k}}\frac{\partial}{\partial R_{\vec
k}}\right)+4
\sum_{\vec\varepsilon}\sin^2\frac{\vec k\vec\varepsilon}{2}
R_{\vec k} R_{-\vec k}\right\} \, .
\end{equation}
The eigensolutions $\psi_K$ of $\hat h_{\rm red}$
in the weak coupling approximation are decoupled harmonic
oscillators
for each $\vec k$, with frequencies
\begin{equation}
\omega^2_{\vec k }=4\sum_{\vec\varepsilon}\sin^2\frac{\vec
k \vec\varepsilon}{2}.
\end{equation}
Because of the `radial Laplacian' it looks as if the eigenfunctions
would have to vanish at the origin to be normalizable. However,
as in the Schr\"odinger equation in three dimensions,
the Jacobian $J$  allows a constant wave function at the origin.
Consequently, the eigenvalue of $\psi_K$
is given by the sum over the modes:
\begin{equation}
\Omega_K=\sum_{\vec
k}\sqrt{4\sum_{\vec\varepsilon}\sin^2
\left(\frac{\vec k\vec\varepsilon}{2}\right)} \, ,
\end{equation}
which gives in the $N_{\bot} \to\infty$ limit spin waves with
$\omega_{\vec k}
=\sqrt{k^2_1+k^2_2}$.

In the weak coupling limit the zero mode Hamiltonian supports
solutions similar to QED. The strong coupling limit, however,
yields different results: `gluonic'
excitations are suppressed because of large energy gaps.
This is due to the Jacobian, which can be traced back
to non-abelian self interactions in the original Lagrangian.

\subsection{Effective Hamiltonian for Two-Site Truncation}

We will now invoke an effective interaction approach embedded in
a cluster expansion~\cite{tjthesis},
starting with the simplest,  two--site cluster, in which either site (or
both) can be excited to high  energy states.  We will obtain the solution
for the low--lying spectra of the system approximated as a low density of
excitable two--site clusters over the entire range of coupling.
This method can be envisaged as the
starting point of more ambitious renormalization group
techniques~\cite{tjf1, MW}.

We work in the representation of the strong coupling solution
of $\hat{h}^{\rm e}_{\vec b} $ and
divide the  two--site subspace into a $P$ and $Q$ space, such that
$P+Q=1$ with
\begin{eqnarray}
P&=&\{|0,0\rangle\}\, ,\nonumber\\
Q&=&\{|n,m\rangle;\ n, m \not= 0,0\}\,  ,
\end{eqnarray}
where $n,m$ represent the indices of the Gegenbauer polynomials.
Then the two--site energy $E_2$ is given by the
non-perturbative solution of the Hamiltonian in Eq. (\ref{hred}),
truncated to two lattice sites and labelled $h_{2}$.
Explicitly, this Hamiltonian is:
\begin{equation}
\hat h_{2}=\hat h^{\rm e} + \hat h^{\rm m},
\end{equation}
with
\begin{equation}
\hat h^{\rm e} = -g^{2}_{\rm eff}\left
\{\frac{1}{J}\frac{\delta}{\delta\varphi_1}
J\frac{\delta}{\delta\varphi_1} +
\frac{1}{J}\frac{\delta}{\delta\varphi_2}
J\frac{\delta}{\delta\varphi_2}
\right \} ,
\end{equation}
and
\begin{equation}
\hat h^{\rm m}=\frac{1}{g^{2}_{\rm eff}} (\varphi_{1}-\varphi_{2})^2,
\end{equation}
where the subscripts label the sites.

Within the effective Hamiltonian
method, the two--site energy is given by~\cite{zvb}:
\begin{equation}
E_2=P\hat h_{2} P+ P\hat h_{2} Q\frac{1}{E_2-Q\hat
h_{2} Q}
Q\hat h_{2} P \, .
\label{blocheq}
\end{equation}
The self--consistent solutions of this
equation provide the low--lying spectra in this method.
The strong coupling basis states are eigenstates of $\hat h^{\rm e}$:
\begin{equation}
\hat h^{\rm e} |n,m \rangle = g^2_{\rm eff}\left \{ n(n+2)+m(m+2)
\right \} |n,m \rangle  \, .
\label{unpertener}
\end{equation}
Thus, the non--trivial matrix elements are those of $\hat h^{\rm m}$,
and are of the form
$\langle n,m | (\varphi_1-\varphi_2)^2 |n',m' \rangle \,$ ,
where the subscripts label the sites.
The matrix element reduces to sums and
products of one--site matrix elements, which are given as:

\begin{eqnarray}
\langle n|\varphi| n'\rangle &=& \frac{\pi}{2} \hspace{4.3cm}{\rm for}\
n=n'  \nonumber \\
\langle n|\varphi| n'\rangle &=&\left\{\begin{array}{cl}
\frac{2}{\pi}\left(\frac{1}{(n+n'+2)^2}-\frac{1}{(n-n')^2}\right)
&{\rm for}\  n+n'= {\rm odd}\\
0&{\rm for}\  n+n'= {\rm even}, n \neq n'
\end{array}\right.  \nonumber\\
\langle n|\varphi^2| n' \rangle&=&\left\{\begin{array}{ll}
\frac{2}{\pi}\left[\frac{\pi^3}{6}-\frac{\pi}{[2(n+1)]^2}\right]
&{\rm for} \ n=n'\\
\frac{2}{\pi}\left\{\pi(-1)^{n+n'}\left[\frac{1}{(n-n')^2}
-\frac{1}{(n+n'+2)^2}\right]\right\}
&{\rm for}\  n\not= n'.\end{array}\right.
\end{eqnarray}

In the strong coupling limit we
obtain the result of perturbation theory in $1/g_{\rm eff}$
\begin{equation}
E_{2}=\langle 0,0|\hat h^{\rm m}|0,0\rangle= \frac{1}{g^2_{\rm eff}} \left \{
\langle 0 | \varphi^2 |0  \rangle  - 2
\left( \langle 0 | \varphi |0  \rangle \right )^2 \right \}
= \frac{1}{g^2_{\rm eff}}
\left(\frac{\pi^2}{6}-1\right).
\label{2pert}
\end{equation}

In the weak coupling limit we can solve the Schr\"odinger equation for
the two--site
version of the earlier spin wave Hamiltonian, Eq. (\ref{spin}).
We call the respective variables $\alpha_{\vec b_1}=x$
and $\alpha_{\vec b_2}=y$ ,
then solve:
\begin{eqnarray}
\hat h_{\rm red}&=&
-\left(\frac{\partial^2}{\partial x^2}
+\frac{2}{x}\frac{\partial}{\partial x}\right)
-\left(\frac{\partial^2}{\partial y^2}
+\frac{2}{y}\frac{\partial}{\partial y}\right) + (x-y)^2.
\end{eqnarray}
Since this Hamiltonian is invariant under $x
\leftrightarrow y$,
the eigenfunctions $\Psi_2(x,y)$
can be chosen to be symmetric under the interchange of $x$ and $y$
($\Psi_{2s}(x,y)$), or antisymmetric
($\Psi_{2a}(x,y)$).
The first symmetric excited state
becomes degenerate with the ground state of the original problem in the
weak coupling limit, and the first antisymmetric state has a greater
energy than the first symmetric state.

As usual one factorizes the wave function
$\Psi_2(x,y)=\frac{1}{xy}\Phi_2(x,y) \,$ ,
resulting in the Schr\"odinger equation
\begin{equation}
\hat h_{\rm red} \Phi_2(x,y) = \left \{ -\frac{\partial^2}{\partial x^2}
-\frac{\partial^2}{\partial y^2} + (x-y)^2 \right \} \Phi_2(x,y) \,.
\end{equation}
The center--of--mass motion is then separated:
$\Phi_2(x,y)=e^{iPR}\chi_2(r) \,$ ,
with $R=(x+y)/2$ and $r=x-y$.
The Hamiltonian corresponding to the relative motion ($r$) is a simple
radial harmonic oscillator:
\begin{equation}
(\hat h_{\rm red})_{r}=-2\frac{\partial^2}{\partial r^2} + r^2.
\label{radial}
\end{equation}
The lowest states of the symmetric and antisymmetric `towers' are solutions
to this Hamiltonian.
The energies of these states can be read directly from Eq.
(\ref{radial});
$ E_{2s}=\sqrt{2}$, and $ E_{2a}=3 \sqrt{2},$
respectively,
giving a energy gap between the states of $2 \sqrt{2}$.

Thus, the results for the energy gaps of the low--lying states in the
weak coupling limit are
\begin{eqnarray}
E_{2s}-E_{\rm ground} &=& 0 \, ,\nonumber \\
E_{2a}-E_{\rm ground} &=& 2 \sqrt{2}  \, .
\label{analgap}
\end{eqnarray}

We now proceed to calculate the low--lying spectra
via the effective Hamiltonian method. In the numerical calculations, we
truncate the $Q$ space at a certain two--site energy, calculate $E_2$,
then increase the energy until we reach convergence
for each choice of coupling constant $g_{\rm eff}$.
The typical number of two--site states required for the $Q$ space
at convergence was 300.

The numerical solution of
Eq. (\ref{blocheq}) for $E_2$ has now been published~\cite{hwln1}.
In the strong coupling limit (large  $g^2_{\rm eff}$) the large gaps in
energy are evident, and the numbers agree with the unperturbed energy of
the states, given in Eq. (\ref{unpertener}).
In this same limit, the slope of the ground state energy
as a function of
the inverse square coupling agrees with the analytic calculation
of Eq. (\ref{2pert}).
In the weak coupling limit, the results for the gap energies were
Richardson extrapolated
for the $1/g^2_{\rm eff} \to \infty$ limit.  This extrapolation
matched the analytical results of Eq. (\ref{analgap}) to five
significant figures.  Thus, we have obtained two--site solutions in the
strong coupling basis for the entire range of coupling which
agree with analytic results in the both the weak and strong coupling
limits.

It is remarkable that we succeeded with a one--dimensional strong
coupling basis throughout the range of coupling strengths.
Results for the spectra of $N$--sites
are straightforwardly obtained and should be valid as long
as the number of
excited two--site clusters is small compared with $N/2$.
This is the `low density' approximation.

\section{Variational Tamm-Dancoff Application}

Over the past several years, we have also investigated
the use of relativistic wave equation formulations of
quantum field theory in the equal time
quantization scheme.  Here, we would like to briefly
summarize our approach as viewed from the
effective Hamiltonian framework.~\cite{sp1,sp2}

We begin by selecting the Coulomb gauge Hamiltonian of QCD and
we choose to work in a finite momentum space domain.  We
eliminate all interactions except the fermion-gluon vertex and
the triple gluon coupling terms.  We fix the scale by giving
the quarks a constituent quark mass.

For our unperturbed Hamiltonian we select the kinetic energy
operators for the quarks and the gluons.  In our earlier
work~\cite{sp1} we chose a cubic B-spline basis expansion
for our quark and gluon amplitudes.  More recently,~\cite{sp2}
we select a harmonic oscillator basis with a variable
oscillator parameter, resulting in a non-orthogonal basis.

For our P-space, we select a small finite set of
quark-antiquark states for mesons and 3-quark
states for baryons.

Our procedure is to optimize our P-space using a variational
approach and to determine an effective interaction for that
basis.  The effective interaction emerges as a ladder series
as given by the G-matrix defined earlier.  However, there are
a number of simplifying assumptions needed to evaluate
this effective interaction.

Finally, we solve the resulting Eq.(\ref{h9}) which is the
Bethe-Salpeter equation with the kernel defined by the
effective interaction we have derived.  For fixed
constituent quark masses and a fixed strong coupling
constant, the solutions of this equation yield
spectra in reasonable accord with data.~\cite{sp1,sp2}

The most remarkable feature of this approach is found in the
shape of the effective interaction which emerges.  At short
distances it closely approximates the one gluon exchange
effective potential and, at large distances, it closely
approximates linear confinement.

The linear confining feature is traced to the non-linear
role of the triple-gluon coupling term and its effect in
our variational treatment.  The combination of the
choice of Coulomb gauge and the variational approach
appear to be ultimately responsible for this result.
We believe that this feature will persist as we make
our approach more general.

\section{Conclusions}

The well established framework of effective Hamiltonians for
quantum many-body theory with strong interactions appears
adaptable to non-perturbative applications in
quantum field theory.  Our initial results on simple
renormalization problems are encouraging and the initial
applications to (simplified) QCD problems also inspire us
to proceed.  However, we realize that considerable
effort is required for major additional advances to be
achieved.

\section*{Acknowledgments}
Three of us
(J.P.V., T.J.F. and J.R.S.) acknowledge partial support by the U.S.
Department of Energy under Grant No. DE-FG02-87ER-40371, Division
of High Energy and Nuclear Physics. One of us (J.P.V.) also
acknowledges the support of the Alexander von Humboldt Foundation.


\section*{References}
\begin{thebibliography}{99}

\bibitem{zh1}D.C. Zheng, B.R. Barrett, J.P. Vary, W.C. Haxton
and C.-L. Song, {{\em Phys. Rev.} C}{\bf 52}2488(1995);
and references therein.

\bibitem{kirson} M.W. Kirson, in {\it Nuclear Shell Models}, eds.~M.
	Vallieres and B.H. Wildenthal (World Scientific, Singapore, 1985),
	p.290.

\bibitem{bloch} C. Bloch and J. Horowitz, Nucl. Phys. {\bf 8} 91(1958).

\bibitem{bran} B.H. Brandow, Rev. Mod. Phys. {\bf 39} 711(1967).

\bibitem{bk1} B.R. Barrett and M.W. Kirson, in
	{\it Advances in Nuclear Physics}, Vol.6, eds.~M. Baranger
	and E. Vogt (Plenum Press, New York, 1973), p.219.

\bibitem{eo} P.J. Ellis and E. Osnes, Rev. Mod. Phys. {\bf 49} 777(1977).

\bibitem{fesh} H. Feshbach, Ann. Phys. (N.Y.) {\bf 19} 287(1962).

\bibitem{ls} S.Y. Lee and K. Suzuki, Phys. Lett. {\bf 91B} 79(1980);
	K. Suzuki and S.Y. Lee, Prog. of Theor. Phys. {\bf 64} 2091(1980).

\bibitem{qbox} T.T.S. Kuo, in {\it Lecture Notes in Physics}, {\bf 144},
	eds.~T.T.S. Kuo and S.S.M. Wong (Springer Verlag, Berlin, 1981), p.248.

\bibitem{zvb} D.C. Zheng, J.P. Vary, and B.R. Barrett, Nucl. Phys. {\bf A560}
 211(1993).

\bibitem{gmatrix} K.A. Brueckner, Phys. Rev. {\bf 97} 1353(1955);
	{\bf 100} 36(1955).

\bibitem{bar1}B.R. Barrett, D.C. Zheng, J.P. Vary, and R.J. McCarthy,
in {\it Recent Progress in Many-Body Theories}, Vol. 4, Ed. by
E. Schachinger {\it et. al}, Plenum Press, New York (1995) p. 163.

\bibitem{wilson} K. G. Wilson and J. Kogut, Phys. Rep. {\bf 12 C}, 75
(1974);
K. G. Wilson, Rev. Mod. Phys. {\bf 47}, 773 (1975).

\bibitem{tjf1}T. J. Fields, J. P. Vary, and K. S. Gupta,
Mod. Phys. Lett. {\bf A11}, 2233(1996).

\bibitem{pinsky} B. van de Sande and S. S. Pinsky, Phys. Rev. {\bf D
46}, 5479
(1992)

\bibitem{huang} {\it Quarks, Leptons and Gauge Fields}, K. Huang
(World Scientific, Singapore (1982)).

\bibitem{hwln1}H.W.L. Naus, H.J. Pirner, T.J. Fields and J.P. Vary,
Phys. Rev. D {\bf56}, 8062(1997).

\bibitem{LNT} F. Lenz, H. W. L. Naus, and M. Thies, Ann. Phys. (N.Y.)
{\bf 233}, 317 (1994).

\bibitem{LMT} F. Lenz, E. J. Moniz, and M. Thies, Ann. Phys. (N.Y.)
{\bf 242}, 429 (1995).

\bibitem{Bronz} J. B. Bronzan, Phys. Rev. D {\bf 37}, 1621 (1988).

\bibitem{tjthesis}T. J. Fields, Ph.D. thesis, Iowa State University, 1996.

\bibitem{Shaba} S. V. Shabanov, Phys. Lett. B {\bf 318}, 323 (1993).

\bibitem{MW} C. J. Morningstar and  M. Weinstein, Phys. Rev. D {\bf 54},
4131 (1996).

\bibitem{zvb} D. C. Zheng, J. P. Vary, and B. R. Barrett, Nucl. Phys.
{\bf A560}, 211 (1993).

\bibitem{sp1}J.R. Spence and J.P. Vary,  {{\em Phys. Rev.} C}
{\bf 52}1668(1995); and references therein.

\bibitem{sp2}J.R. Spence and J.P. Vary, `Variational
Tamm-Dancoff Treatment of Quantum Chromodynamics II:
A Semi-Analytic Treatment of the Hadrons in the Valence
Quark Approximation,' to be published

\end{thebibliography}
\end{document}